\documentclass[aps,prl,twocolumn,showpacs,groupedaddress]{revtex4}  
\usepackage{graphicx}  
\usepackage{dcolumn}   
\usepackage{bm}        
\usepackage{amssymb}   
\usepackage{ulem}
\usepackage{epsfig} 
\usepackage{array} 
\usepackage{amsmath} 
\usepackage{lscape}
\usepackage{color}
\usepackage{graphicx}
\newcommand{\bal}[1]{\textcolor{black}{#1}}

\begin{document}
\title{Impact of  impellers on the axisymmetric magnetic mode in the VKS2 dynamo experiment}
\author{R. Laguerre,$^{1,2}$ C. Nore,$^2$
 A. Ribeiro,$^2$ J. L\'eorat,$^3$ J.-L. Guermond$^4$
and F. Plunian$^5$} 

\email{nore@limsi.fr}

\affiliation{$^1$ Universit\'e Libre de Bruxelles, CP.231, Boulevard
  du Triomphe, Brussels, 1050, Belgium; $^2$Laboratoire d'Informatique
  pour la M\'ecanique et les Sciences de l'Ing\'enieur, CNRS,
  Universit\'e Paris-Sud 11 BP 133, 91403 Orsay cedex, France;
  $^3$Luth, Observatoire de Paris-Meudon, place Janssen, 92195-Meudon,
  France; $^4$Department of Mathematics, Texas A\&M University,
  College Station, TX 77843, USA; $^5$ Universit\'e Joseph
  Fourier, CNRS, Laboratoire de G\'eophysique Interne et de
  Tectonophysique, 38041 Grenoble, France} \date{\today}
\begin{abstract}
  In the VKS2 (von K\'arm\'an Sodium 2) successful dynamo experiment of September 2006, the
  magnetic field that was observed showed a strong axisymmetric
  component, implying that non axisymmetric components of the flow field
  were acting.  By modeling the induction effect of the spiraling flow
  between the blades of the impellers in a kinematic dynamo code, we
  find that the axisymmetric magnetic mode is excited and becomes
  dominant in the vicinity of the dynamo threshold.
  The control parameters are the magnetic Reynolds number \bal{of the mean flow}, the
  coefficient measuring the induction effect, $\alpha$, and the type
  of boundary conditions (vacuum for steel impellers and normal field
  for soft iron impellers). We show that using realistic values of
  $\alpha$, the observed critical magnetic Reynolds number, $Rm^c
  \approx 32$, can be reached easily with ferromagnetic boundary
  conditions. We conjecture that the dynamo action achieved in this
  experiment \bal{may not be} related to the turbulence in the bulk of the flow,
  but 
  \bal{rather to} the alpha effect \bal{induced by} the impellers.
\end{abstract}
\maketitle

\bal{The interest of the scientific community for the dynamo action 
in liquid metal has been renewed since 2000 in the wake of successful
experiments~\cite{Ga2000,StMu01,Monchaux06}. We focus in this Letter on
the Cadarache VKS2 (von K\'arm\'an Sodium 2) experiment~\cite{Monchaux06} where the dynamo
effect occurred beyond the critical magnetic Reynolds number $Rm^c
\approx 32$.}
\bal{The growing magnetic field that was observed
above threshold was mainly axisymmetric.
Dynamo action was found with soft iron impellers but did not occur with
stainless steel impellers, using the same available power.}  
\bal{The purpose of the present Letter is to present arguments based on
modeling and numerical simulations to justify the observed axisymmetric mode.
The two key ingredients are the so-called
$\alpha$-effect and ferromagnetic boundary conditions.}

From Cowling's theorem
\cite{Cowling34} the axisymmetric part of the flow cannot be
responsible alone for the generation \bal{of an} axisymmetric field.
\bal{In the experiment, the} deviation from axisymmetry of the velocity field is thus expected
to participate to the dynamo process, generating an axisymmetric
electromotive force somewhere in the fluid and triggering the dynamo
growth.  Two other magnetohydrodynamic experiments using
counter-rotating propellers have also led to this conclusion
either on observational~\cite{Lathrop00,Forest06} or on numerical
grounds~\cite{Forest07}. It was argued in
\cite{Lathrop00,Forest06,Forest07} that turbulence effects were
responsible for the occurrence of the axisymmetric magnetic field. \bal{
In~\cite{Petrelis07} the turbulence effects were held responsible for the
dynamo threshold in VKS2. In addition an $\alpha$-effect close to the discs was
invoked to justify the axisymmetry of the generated field.} 
 
\begin{figure}
\epsfig{file=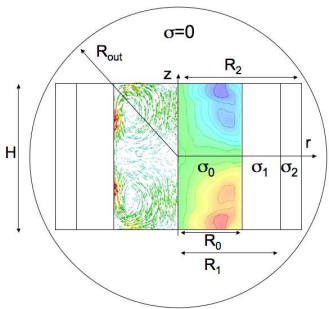,width=0.3\textwidth}
\caption{Geometry of the computational domain in a meridional plane.
  The axisymmetric meridional flow $(U_r,U_z)$ is represented on the
  left with arrows. On the right the isovalues of $U_{\theta}$ are
  plotted. The radii $R_i$ and height $H$ as well as the
  conductivities $\sigma_i$ are defined in the text.}
\label{meanflow}
\end{figure}

The VKS2 configuration is modeled as follows: The computational domain
is divided into three concentric cylinders of axis $z$, embedded in an
insulating sphere (figure~\ref{meanflow}).  The inner cylinder $r\le
R_0$ contains the moving sodium of conductivity $\sigma_0$.  The
cylindrical shell $R_0 \le r \le R_1$ contains stagnant sodium
($\sigma_1=\sigma_0$). The cylindrical shell $R_1 \le r \le R_2$ is a
layer of copper ($\sigma_2=4.5 \sigma_0$).  We choose $R_0 $ as unit
length and the geometric parameters are $R_1/R_0=1.4$, $R_2/R_0=1.6$
and $R_{out}/R_0=10$, with the same height $H/R_0=1.8$ for the three
cylinders.
The velocity field of the liquid sodium $\textbf{U}$ ($r<R_0$) is
axisymmetric and is obtained from time averaged measurements of a
water experiment in the VKS2 setting~\cite{Ravelet05}.  The meridional
flow \bal{field and the isolines of the azimuthal component of the velocity}
are shown in figure~\ref{meanflow}.
The flow field is maintained by two counter-rotating impellers 
located at both ends of the vessel.
They each occupy a cylindrical volume of radius $R_0 $ and thickness $H/20$.

\begin{figure}
\includegraphics[width=0.4\textwidth]{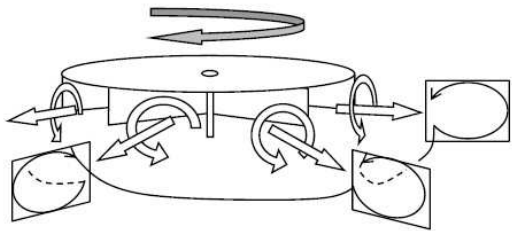} \\*[0cm]
\caption{Sketch of the top impeller with eight radial blades attached
  to the disc. Due to the rotation of the disc (gray arrow on the
  top), radial helical vortices are produced between the blades (thick
  white arrows). An axisymmetric azimuthal field
  $(\textbf{B}\cdot\textbf{e}_{\theta}) \textbf{e}_{\theta}$ (thin
  line) is distorted leading to eight magnetic loops.}
\label{blades}
\end{figure}

The only non axisymmetric parts of the device are the blades composing
the impellers (eight each), and we represent the effect of these blades
by adding the electromotive force $\alpha
(\textbf{B}\cdot\textbf{e}_{\theta}) \textbf{e}_{\theta}$ to the
induction equation. This force is assumed to act only in the two fluid
cylinders occupied by the blades as they rotate.  This model is
deduced as follows: As the blades rotate, the fluid is ejected
radially outwards and radial vortices are created due to the rotation
rate gradient, see the white arrows in figure~\ref{blades}.
We
then assume that when a magnetic field is applied, the mean induced
current is dominated by its azimuthal component and that the shearing
action of the vortices is cumulative. The axial current is expected to
vanish in the vicinity of the discs \bal{and, as a result, we set the $\alpha_{zz}$
component of the $\alpha$ tensor to be zero}. This phenomenon is illustrated on
fig.~\ref{blades}. 
As the magnetic lines
$(\textbf{B}\cdot\textbf{e}_{\theta}) \textbf{e}_{\theta}$ (thin line
in figure~\ref{blades}) are distorted by the helical vortices, small
scale magnetic loops are created, which in turn create an azimuthal
electromotive force $\alpha
(\textbf{B}\cdot\textbf{e}_{\theta})\textbf{e}_{\theta}$, where the
parameter $\alpha$ is negative and depends on the impeller rotation
rate (see below for an estimate), the size, curvature and number of
blades. We thus attempt to represent the complexity of the induction
effects in a poorly known flow close to a boundary by a single scalar
parameter.  The handedness of the helical vortices generated between
the blades is directly linked to the rotation sign
of the impellers: orienting the rotation axis of the bottom impeller
towards the center of the container, a positive rotation
drives a set of eight radial positive
helices creating a negative $\alpha$ coefficient.
\bal{
The counter-rotation configuration with $\Omega_{\text{bot}}=
-\Omega_{\text{top}}$ is invariant by rotation of $\pi$ about any horizontal
axis~\cite{NTDX03}, therefore the two impellers produce the same
$\alpha$. Actually, whatever} \bal{the sign for the rotation of the bottom impeller, $\Omega$,
the product $\Omega \alpha$ is always negative. Henceforth we set $\Omega_{\text{bot}}=\Omega >0$. }

We follow the standard kinematic dynamo approach by solving the
induction equation
\begin{equation}
  \partial_t \textbf{B} = \nabla \times (\widetilde{\textbf{U}} \times \textbf{B} 
  + \alpha (\textbf{B}\cdot\textbf{e}_{\theta}) \textbf{e}_{\theta}) 
  - \nabla {\times} (\eta \nabla{\times}\textbf{B}).
\label{inducequation}
\end{equation}
where $\eta = 1/\sigma\mu_0$, $\mu_0$ being the vacuum magnetic
permeability, $\sigma$ the fluid conductivity and $\alpha$ the unknown
parameter which models the induction effect in the volume occupied in
average by the impellers.  The field $\widetilde{\textbf{U}}$ is equal
to $\textbf{U}$ in the cylinder $ r\le R_0$ and is zero in the two
cylindrical shells $R_0\le r\le R_1$, $R_1\le r\le R_2$. The
conductivity field is such that $\sigma=\sigma_0$ in the inner
cylinder $ r\le R_0$ and in the stagnant sodium shell, and
$\sigma=\sigma_2$ in the copper layer.  The conductivity jump at
$r=R_1$ is accounted for by enforcing the continuity of the tangent
component of the electric field.  Note that the flow behind the
impellers has not been simulated (see discussion below). To simplify
the parameterization of the alpha effect, $\alpha$ vanishes outside two
thin cylindrical layers of radius $R_0 $ and thickness $H/20$ situated at each end
of the container and its intensity varies smoothly within these
layers \bal{using a $\tanh$ regularization function}.  
The magnetic Reynolds number is $Rm \equiv \sigma_0 \mu_0
U_0R_0$ where $U_0$ is \bal{the} maximum speed of the flow $\textbf{U}$.

Two types of boundary conditions are compared to model the
experiment: they are referred to as insulating (I) and ferromagnetic
(F) boundary conditions.  We say that insulating boundary conditions
(I) are used when the continuity of the magnetic field is enforced at
the interface between the conducting regions and the vacuum (i.e., at
${|z|=H/2, \, 0\le r \le R_2}$ and at ${0 \le |z| \le H/2, \,
  r=R_2}$). We say that ferromagnetic boundary conditions (F) are used
when we enforce $\textbf{B}\times \textbf{e}_z=0$ at ${|z|=H/2, \,
  0\le r \le R_0}$ and we require $\textbf{B}$ to be continuous across
the other boundaries ${0 \le |z| \le H/2, \, r=R_2}$ and ${|z|=H/2, \, R_0\le r \le R_2}$.  The condition
$\textbf{B}\times \textbf{e}_z=0$ is meant to mimic the effect of
ferromagnetic discs with infinite magnetic permeability located at the
top and bottom of the vessel\bal{~\cite{Durand68}}.  These boundary conditions model steel
impellers (I) and soft iron impellers (F), respectively.

\begin{figure}
  \includegraphics[width=0.4\textwidth]{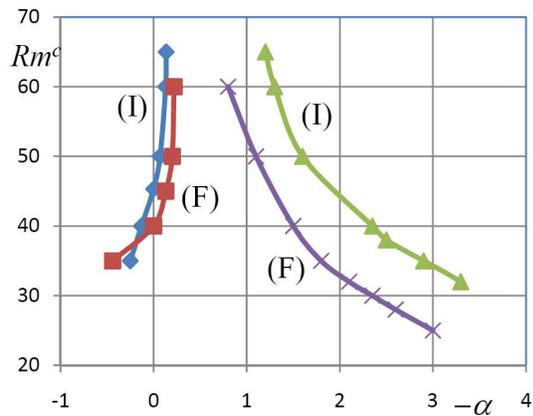} \\*[0cm]
  \caption{Dynamo threshold $Rm^c$ versus $-\alpha$, for ($\times$,
    $\triangle$) $m=0$, ($\square$,$\diamond$) $m=1$ and where (F) and
    (I) indicate ferromagnetic and insulating boundary conditions
    respectively.}
\label{marginal}
\end{figure}

Our numerical code~\cite{Laguerre06, Guermond07} uses a finite element
Galerkin method for the spatial discretization in the meridional plane
$(r,z)$ and a Fourier series decomposition along the azimuthal
direction.  Since the coefficients of the induction equation do not
depend on the azimuthal angle, the different azimuthal modes
($m=0,1,2, \ldots$) are decoupled. The time is discretized with a
semi-implicit backward finite difference method of second order.
Growth rates are computed for the two types of magnetic boundary
conditions, (I) or (F), and for various values of $\alpha$ and $Rm$. The
zero growth rate condition defines a critical curve in the ($\alpha$,
$Rm$) plane. A compilation of results obtained for the (I) and (F)
cases and for Fourier modes $m=0$ and $m=1$ is shown in figure~\ref{marginal}.

\begin{figure}
  \begin{center}
\epsfig{file=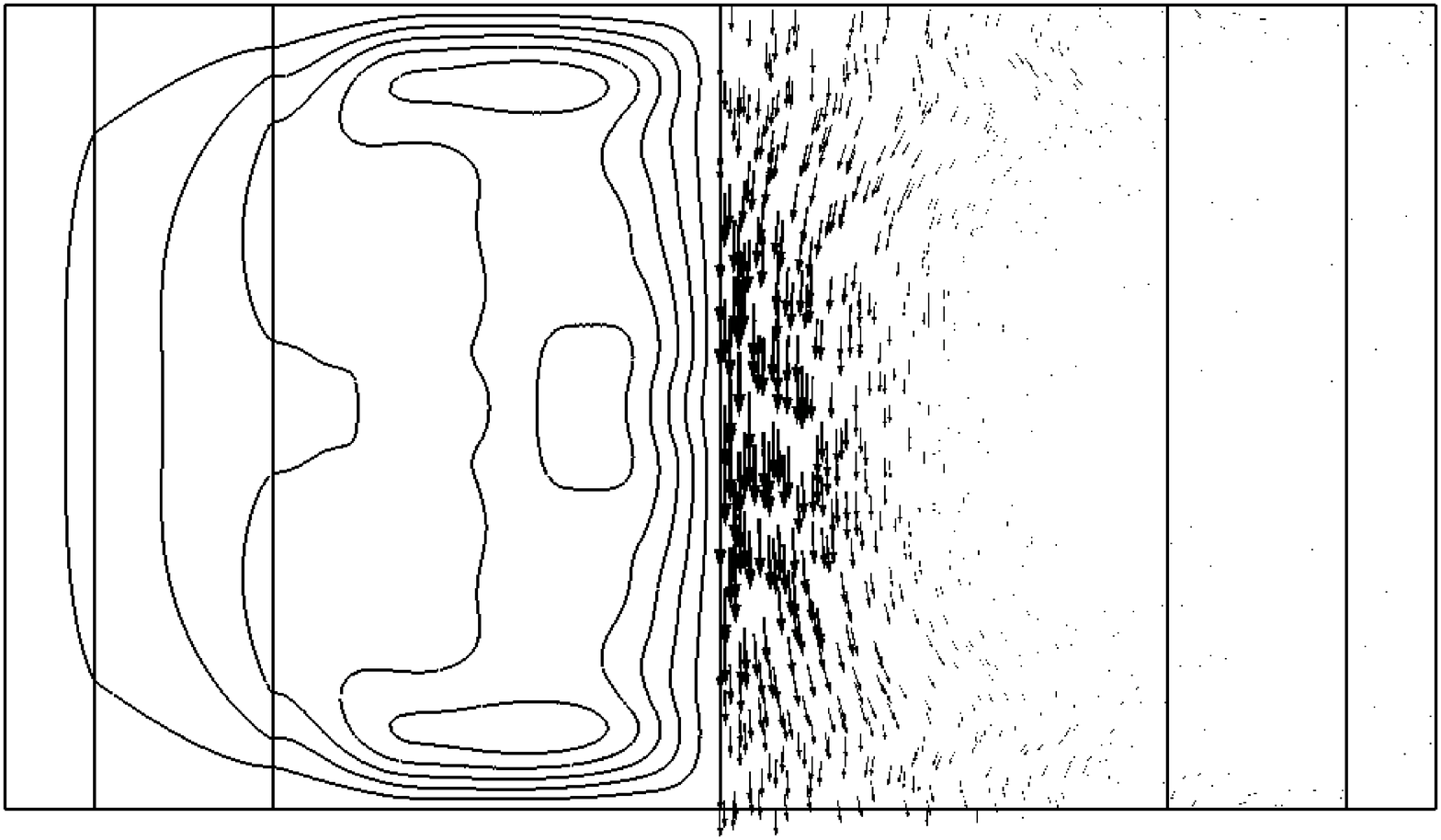,width=0.4\textwidth}
\epsfig{file=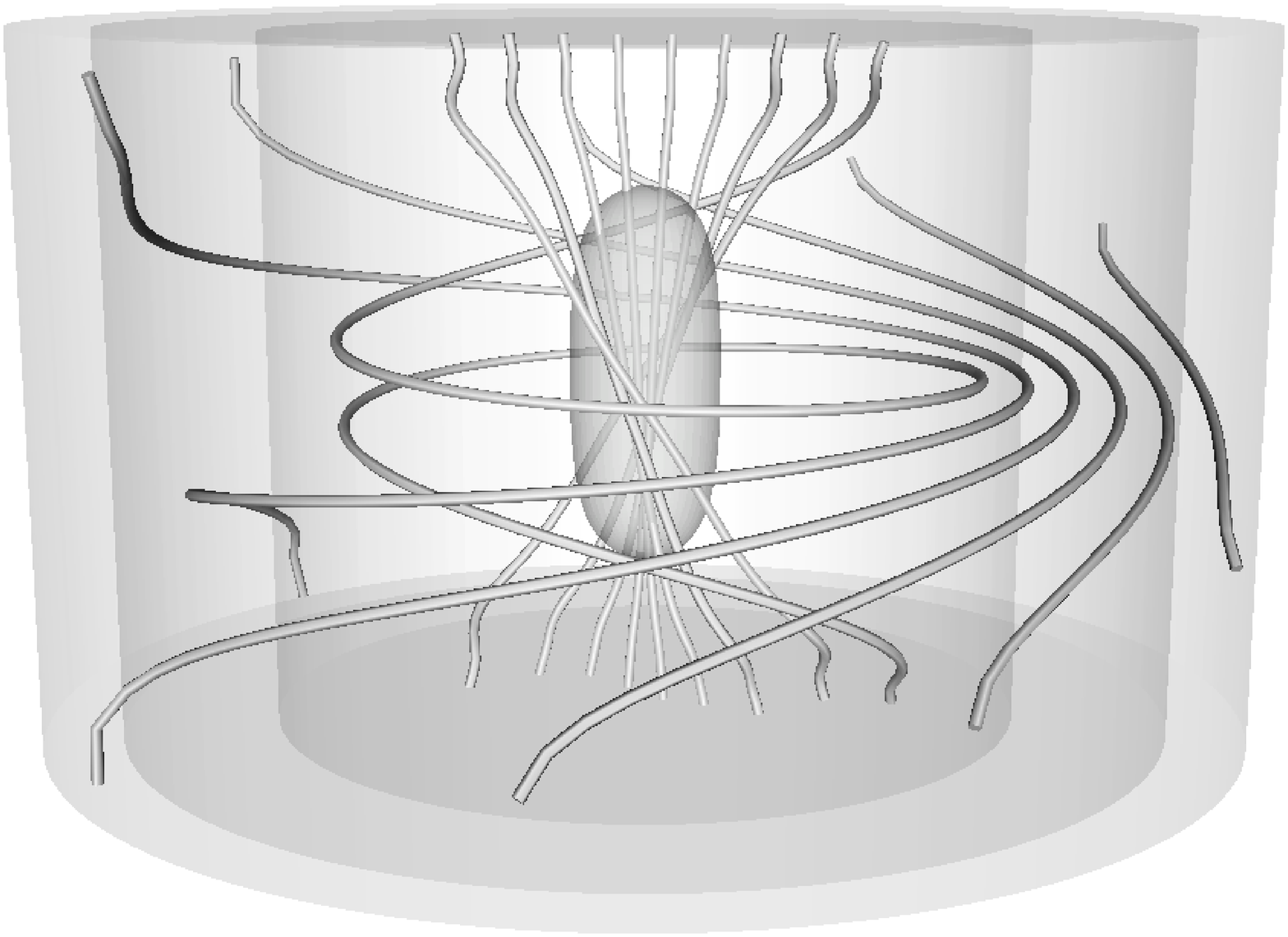,width=0.45\textwidth}
  \caption{Top: Geometry of the $m=0$ (F) growing magnetic field for
    $Rm^c=32$, $|\alpha|=2.2 > |\alpha_c|=2.1$ . The three cylinders
    are also represented.  The lines correspond to the positive
    isovalues of $B_{\theta}$ and the arrows to the meridional field
    $(B_r,B_z)$. Bottom: isovalue of 25\% of the maximum energy
    plotted in the 3D domain, together with some field lines.}
\label{field}
\end{center}
\end{figure}
In order to make comparisons with experimental results published
in~\cite{Monchaux06}, we first focus our attention on the axisymmetric
magnetic mode ($m=0$), which was a priori eliminated in former studies
by Cowling's theorem.
\bal{The two critical curves corresponding to the boundary
conditions (I) and (F) and the Fourier mode $m=0$ reveal
that the dynamo action occurs only for $\alpha$ negative,
thus confirming the phenomenological
argument discussed above to justify the sign of $\alpha$.} 
Both curves clearly show that
increasing $|\alpha|$ lowers the critical magnetic Reynolds
number. Moreover, the (F) boundary condition yields significantly
smaller thresholds. This observation agrees with previous results
obtained in a different context~\cite{Avalos03, Avalos05}. Both curves
can be represented by the scaling law $Rm^c \sim |\alpha|^{-q}$, where
$q \approx 0.67$, suggesting that the dynamo process is similar for
both types of boundary conditions.

The value of $\alpha$ corresponding to the dynamo threshold in VKS2,
$Rm^c=32$, is found to be $|\alpha|=2.1$ for the (F) boundary condition and
$|\alpha|=3.3 $ for the (I) boundary condition.  Since the steel impellers have not
produced dynamo action, this suggests that, as a first guess, the
effective $|\alpha|$ for VKS2 is between 2.1 and 3.3.  A rough
estimate of $\alpha$ is given by $\alpha \sim u^2 h /\eta$ where $u$
is a typical flow intensity between the blades and $h=H/20$. From
\cite{Ravelet05} we estimate $u/U_0\sim 0.2$ at the impeller half
radius. Then, for $Rm=32$, $\eta$=0.1 m$^2$.s$^{-1}$ and $R_0=0.2$m,
we find $\alpha \sim 1.8$m.s$^{-1}$, making the value $|\alpha|$ =
2.1m.s$^{-1}$ plausible.  If $|\alpha|$ is indeed close to 2.1, the critical
value of $Rm$ for the (I) boundary condition is close to 43, which
apparently contradicts the experiments since $Rm= 50$ has been reached
without dynamo action using steel impellers.  To explain this
discrepancy we recall that it has been shown in \cite{Stefani06} that
for the Fourier mode $m=1$ the flow of liquid sodium behind the
impellers acts against the dynamo action. We expect this anti-dynamo
effect to be active also for the Fourier mode $m=0$, but we did not
include this extra layer of fluid in the present study.
The negative consequences of this secondary flow are probably screened
in the case of soft iron impellers. The numerical evidence justifying this
assertion has to await the availability of a code describing
conducting domains with different magnetic permeabilities (work in progress).

The stationary eigenvector corresponding to the (F) boundary condition
for $m=0$ and $|\alpha|=2.2$ is shown in figure~\ref{field}.  The
magnetic field is mainly concentrated along the $z$-axis in the fluid
region $r\le R_0$, and it is azimuthally dominated 
in the plane $z=0$ 
for $r\ge R_0$. The radial component is odd with
respect to $z$ whereas the azimuthal and vertical components are even
and of opposite sign. These features are compatible with the 
magnetic field measured at saturation in the experimental dynamo regime
obtained with soft iron impellers~\cite{Monchaux06}.
The eigenvector corresponding to the (I) boundary condition
for $m=0$ and $|\alpha|=2.2$ (not shown here)
is associated to an eigenvalue with a nonzero imaginary part. The resulting dynamo is periodic
with a reversing dipolar moment. This result, not detailed
here, is in strong contrast with the case $\alpha=0$ (I), for which
a non oscillatory transverse dipole mode ($m=1$) with
internal banana-like structures is the only 
unstable mode~\cite{Ravelet05}.

We have also investigated the effect of the $\alpha$ model on the
Fourier mode $m=1$. This mode has not been observed in the VKS2
experiment although it has been predicted to be the most unstable in
axisymmetric kinematic dynamo
simulations~\cite{Ravelet05,Marie06,Stefani06,Laguerre_thesis}.
\bal{In figure~\ref{marginal}, for the mode $m=1$, we see that} the critical magnetic Reynolds
number grows sharply with $|\alpha|$ for $\alpha <0$ and
decreases with $|\alpha|$ for $\alpha >0$. We are thus led to
conclude that the growth of the $m=1$ mode is hindered by the
very induction effect that generates the $m=0$ mode!
This surprising result may explain the absence
of the $m=1$ mode in the experiment, at least for $Rm < 50$.

The present study shows that the positive results of the dynamo
experiment of September 2006 (VKS2) as well as the former negative
results (VKS1) may be explained by a simple mean induction effect
induced by the radial vortices trapped between the blades of the impellers.
\bal{We conjecture that turbulence may not play a role as essential as 
initially believed in the VKS experiments.}

\bal{This work also shows the importance of
using ferromagnetic materials, corresponding to the (F) boundary conditions in fig.~\ref{marginal}.
A research program aiming at exploring the saturation regime
using nonlinear computations and materials with
heterogeneous magnetic permeabilities is currently engaged.}

This work was supported by ANR project no. 06-BLAN-0363-01
``HiSpeedPIV''.  We are pleased to acknowledge the Saclay VKS-team for
providing us with a mean velocity field measured in a water
experiment.  We acknowledge fruitful discussions with F.~Stefani.  We
warmly thank D. Carati and B. Knaepen, the organizers of the MHD
Summer Program, Bruxelles, July 2007. The computations were carried
out on the IBM Power 4 computer of IDRIS of CNRS (project \# 0254).

\end{document}